\documentclass[11pt]{article}
\usepackage{rotating}

\usepackage{lscape}

\usepackage{graphicx}
\begin{document}
\title{\bf The Chemical Composition of Globular Clusters of 
Different Nature in our Galaxy }

\author{{V.\,A.~Marsakov,  V.\,V.~Koval', M.\,L.~Gozha,}\\
%EndAName
{Southern Federal University, Rostov-on-Don, Russia}\\
{e-mail:  marsakov@sfedu.ru, litlevera@rambler.ru, gozha\_marina@mail.ru}}
\date{accepted \ 2019, Astronomy Reports, Vol. 63, No. 4, pp. 274-288}

\maketitle

\begin {abstract}

A catalog of Galactic globular clusters 
has been compiled and used  to analyze relations 
between the chemical and kinematic parameters 
of the clusters. The catalog contains positions, distances,
luminosities, metallicites, and horizontal-branch morphology 
indices for 157~globular clusters, as well as space velocities 
for 72~globular clusters. For 69~globular clusters, these data 
are suppleented with the relative abundances of 28~chemical 
elements produced in various nuclear-synthesis processes, taken
from 101~papers published between 1986 and 2018. The tendency 
for redder horizontal branches in low-metallicity accreted 
globular clusters is discussed. The discrepancy between the 
criteria for cluster membership in the thick-disk and halo 
subsystems based on chemical and kinematic properties is 
considered. This is manifest through the fact that all 
metal-rich ($\rm{[Fe/H]} > -1.0$) clusters are located close 
to the center and plane of the Galaxy, regardless of 
their kinematic membership in particular Galaxy subsystems.
An exception is three accreted clusters lost by a dwarf 
galaxy in Sagittarius. At the same time, the
fraction of more distant clusters is high among metal-poorer 
clusters in any kinematically selected Galactic subsystem. 
In addition, all metal-rich clusters whose origins are 
related to the same protogalactic cloud are located in 
the [Fe/H]–[$\alpha$/Fe] diagram considerably higher 
than the strip populated with field stars. All metal-poor 
clusters (most of them accreted) populate the entire 
width of the strip formed by high-velocity (i.\,e., 
presumably accreted) field stars. Stars of dwarf 
satellite galaxies (all of them being metal-poor)
are located in this diagram much lower than accreted 
field stars. These facts suggest that all stellar objects
in the accreted halo are remnants of galaxies with higher 
masses than those in the current environment of the Galaxy. 
Differences in the relative abundances of $\alpha$-process elements 
among stellar objects of the Galaxy and surrounding dwarf 
satellite galaxies confirm that the latter have left 
no appreciable stellar traces in the Galaxy, with the 
possible exception of the low-metallicity cluster 
Rup~106, which has low relative abundances 
of $\alpha$-process elements.

\end{abstract}


\maketitle

%%%%%%INTRODUCTION%%%%%%%%%%
\section{Introduction}

As the oldest objects in the Galaxy, globular clusters
are of considerable interest, since they can be
used to study the formation and early evolution of the
Milky Way. Until recently, all globular clusters were
believed to be typical representatives of the Galactic
halo; i.\,e., they were considered objects formed from a
single protogalactic cloud during early stages in the
formation of the Galaxy. It was later discovered that
some stellar objects had been captured by the Galaxy
from disrupted companion galaxies.

According to current views, high-mass galaxies
such as the Milky Way are formed at early stages
of their evolution as a result of the continuous accretion
of dwarf galaxies. Some of these galaxies
contained globular clusters, which later become bona
fide members of the Galaxy. Numerical simulations of
this scenario shows that accreted clusters dominate
in the Galaxy beyond a Galactocentric distance of
15--20~kpc \cite{1}. The epoch of large-scale accretion of
extragalactic objects probably occurred in the earliest
stages of the Galaxy’s evolution, but individual accretion
events are ongoing even now. In particular,
we are currently observing the disruption of a dwarf
galaxy in the constellation Sagittarius (Sgr) due to
tidal forces exerted by the Milky Way \cite{2,3}. As
was demonstrated in \cite{4}, five globular clusters are
reliably associated, spatially and kinematically, with
this dwarf galaxy: M~54, Arp~2, Ter~8, Whiting~1,
and NGC~5634. Four more clusters belong to the
Sagittarius system with somewhat lower probabilities:
Berkeley~29 (an open cluster), NGC 5053,
Pal~12, and Ter~7; two clusters (NGC~4147 and Pal~2)
can be associated with it with fairly low probabilities.
According to \cite{4}, the clusters M~53, NGC~288, Pal~5,
and Pal 15 probably do not belong to this system,
although the earlier papers \cite{5,6} claimed the opposite.

Tang et\,al. \cite{7} reconstructed the orbit of the low-
metallicity cluster NGC~5053 based on proper motions
derived from 11~years of data obtained with
the Hubble Space Telescope, and rejected a possible
physical association between this cluster and
this dwarf galaxy. (In addition, according to \cite{5,6},
the clusters M~2, M~5, NGC~5824, NGC~6356,
NGC~6426, and Ter~3 may also belong to this system,
but this was subsequently not confirmed.)

It is usually believed that the core of the system
is the very massive globular cluster M~54 \cite{8}.
Massari et\,al. \cite{9} studied the orbit of the globular
cluster NGC~2419 and concluded that it had also
been lost by the Sgr dwarf galaxy. The elements of
the Galactic orbits of the clusters Rup~106, Pal~13,
NGC~5466, NGC~6934, and NGC~7006 also indicate
that they were most likely captured from various
satellite galaxies \cite{10,11}. NGC~1851, NGC 1904,
NGC 2298, NGC 2808, AM2, and Tom~2 were found
in \cite{12} to be reliably associated with the Canis Major
(CMa) dwarf galaxy, while the clusters NGC~4590,
Pal~1, and Rup 106 are considered to be only possible
members of this galaxy. Freeman \cite{13} suggests that
even $\omega$~Centauri ($\omega$~Cen), the largest known globular
cluster in our Galaxy, which is located fairly close to
the Galactic center and has a retrograde orbit, was at
some time the core of a dwarf galaxy. The numerical
simulations of Tshuchiya et\,al. \cite{14} demonstrated that
the disruption of a dwarf satellite by Galactic tidal
forces and the subsequent motion of its central cluster
in the Galaxy along a very elongated orbit with a small
apogalactic radius is quite possible.

Accreted clusters can be reliably identified only by
analyzing their positions and Galactic orbits. However,
the total velocities of some clusters are not yet
known, especially distant ones. It is often assumed
that all clusters at distances from the Galactic center
exceeding 15~kpc have been accreted. For this reason,
they are often identified as a special subsystem, called
the ``outer halo''. In particular, the dual nature of the
inner and outer Galactic halos was demonstrated in
\cite{15}: the outer halo contains more objects in retrograde
orbits, suggesting they are associated with the
accretion of low-mass galaxies. On the other hand,
another pattern was also found: halo clusters possessing
abnormally red horizontal branches, atypical
for their low metallicities, are predominantly located
outside the solar circle, while clusters with extremely
blue horizontal branches are concentrated inside the
solar circle \cite{16}. (Note that an excess number of
stars in the red part of the horizontal branch could
be due to a younger cluster age, so these clusters
were initially taken to be ``young''.) This difference
was explained by suggesting that the subsystem of
old clusters formed together with the Galaxy as a
whole, while the younger clusters were captured by
the Galaxy from intergalactic space at later stages
of its evolution \cite{17}. Precisely these characteristics
were used to distinguish them in order to study them
in more detail. Such clusters are usually assumed
to form a subsystem called the ``young halo'', ``outer
halo'', or ``accreted halo'' (e.\,g., \cite{18} and references
therein). The remaining, genetically related, clusters
are also subdivided into two subsystems: the
thick disk and the halo itself. This subdivision was
motivated by the shape of the clusters' metallicity
distribution, which displays a deep minimum near
$\rm{[Fe/H]} \approx -1.0$ (e.\,g., \cite{18, 19}). In this subdivision,
metal-richer clusters are considered to be representatives
of the Galactic thick disk. Detailed studies of the
positions and orbital elements for metal-rich clusters
demonstrate that most of them instead belong to the
Galactic bulge (see, e.\,g., \cite{20}). Only the remaining
globular clusters can be considered typical representatives
of the Galactic halo proper.

It was long supposed that all the stars in a given
cluster formed simultaneously, so that the elemental
abundances in their stars should correspond to those
in the primary protoclouds from which these clusters
formed. It was later established that all clusters
undergo self-enrichment, which changes the abundances
of some elements (e.\,g., \cite{21} and references
therein). At least two of the highest-mass clusters,
$\omega$~Cen and M~54, are known to have even been enriched
with elements ejected in supernova outbursts,
resulting in the formation of a younger stellar population
in these clusters with higher abundances of
iron-peak elements. However, super-massive clusters
are not numerous, and the remaining clusters
display distorted abundances only of elements participating
in proton-capture reactions taking place in
hydrostatic helium-burning processes at the centers
or in shell sources in stars on the asymptotic giant
branch (AGB). These processes mainly result in reduced
abundances of the primary $\alpha$-process elements
(oxygen and, to a lesser extent, magnesium) and
enhanced abundances of sodium and aluminum in
AGB stars. When such stars eject their envelopes
at later evolution stages, these elements enrich the
cluster's interstellar medium, so that new generations
of cluster stars have altered chemical composition.
The mean abundances of the other elements in cluster
stars remain essentially equal to their initial values
(see, e.\,g., \cite{22} and references therein). Thus, we
can use these abundances to reconstruct the Galaxy's
evolution at early stages of its formation. Globular
clusters belonging to different subsystems were
formed from interstellar matter with different histories
of its chemical evolution, leading us to expect the relative
elemental abundances in clusters with different
natures to also be different.

This paper presents a comparative analysis of the
interrelations between the relative abundances of $\alpha$-
process elements and the spatial and kinematic characteristics
of globular clusters belonging to different
Galactic subsystems, in order to reveal their nature.
For this purpose, we have compiled a catalog
containing the relative abundances of elements
produced in various nucleosynthesis processes taken
from the literature, as well as the spatial-velocity
components and horizontal-branch morphology indices
of as many Galactic globular clusters as possible.

%%%%%% Data %%%%%%%%%%

\section {INPUT DATA}

Our catalog is based on the latest (2010) version
of the catalog compiled by Harris \cite{23}, which includes
measured parameters for 157 Galactic globular clusters.
We supplemented these data with the relative
abundances of 28~elements in stars of 69~globular
clusters taken from 101 papers published in 1986--
2018. In almost all the sources we used, the elemental
abundances in cluster stars were determined from
high-resolution spectra, mainly for red-giant atmospheres,
which were analyzed assuming LTE. For our
study, we compiled previously published abundances
for $\alpha$-process elements (O, Mg, Si, Ca, and Ti), C,
iron-peak elements (Sc, V, Cr, Mn, Fe, Co, Ni, Cu,
Zn), slow and rapid neutron-capture elements (Sr, Y,
Zr, Mo, Ba, La, Ce, Nd, Eu, Dy), and three elements
with odd numbers of protons (Na, Al, K). We
found the cluster [el/Fe] ratios using the [el/H] and
[Fe/H] abundances from the original publications,
rather than from the mean metallicities presented in
this paper.

The mean number of cluster stars studied in a
single paper is 12, and the most probable number
is 10. We did not consider papers with the largest
numbers of cluster stars studied, in the clusters $\omega$~Cen
(855 stars), NGC~104 (181), and NGC~2808 (123).
In several papers, elemental abundances in clusters
were measured based on one or two stars (there
are 46 such determinations), but we were unable to
find abundances for these clusters in other publications
in only eight cases. Abundances based on one
star were the only ones available for NGC~2419 and
Pal~6, abundances based on two stars for NGC~5024,
NGC~5897, NGC~6362, M~71, Arp~2, and Rup~106,
and abundances based on three stars for NGC~5466
and NGC~6352, in all cases with these abundances
published in only one paper. Abundances were presented
in several publications for 30 of the 69 clusters.
For example, there are 2334 elemental abundances for
stars in $\omega$~Cen published in five papers, and 232 abundances
for stars in NGC~104 in nine papers. Having
large numbers of studied stars per cluster considerably
improves the accuracy of the mean elemental
abundances derived for these clusters. The metallicities
and relative elemental abundances in globular
clusters, averaged over all stars, are presented in
Table~1. We have reproduced here only part of Table~1,
which is available in its complete form electronically.
If the abundance of an element in a cluster was determined
in several papers, we calculated weighted mean
values with weights proportional to the numbers of
cluster stars analyzed in each paper. Unfortunately,
because the information needed to reduce all the measurements
to a uniform scale is not available in every
paper, we used all the published measurements with
no changes.

Thanks to the fact that the abundances for some
clusters were measured in more than one paper, we
were able to estimate the external uncertainties. This
was done by analyzing the distributions of the deviations
of the relative abundances in a cluster determined
in each study from their weighted mean values.
The mean dispersions of these distributions (reflecting
the external agreement between mesurements
from different studies) are in the range 
$\langle\sigma\rm{[el/Fe]}\rangle =(0.06-0.16)$. 
The external errors for clusters with
relative elemental abundances determined in several
studies are presented in Table~1; the same column
of this table also presents the abundance dispersions
for cluster stars claimed in the data sources. Such
clusters have only one reference in the last column
in Table~1. Our comparative analysis
shows that, on average, the external uncertainties for
elemental abundances in globular clusters are only
slightly higher than the abundance dispersions for
cluster stars reported for the original data. This
indicates that there are no significant discrepancies
in the abundances determined in different studies,
making it possible to use our compiled abundances
for a statistical analysis of the abundances in clusters
belonging to different Galactic subsystems.

We calculated the Cartesian coordinates of all
157 globular clusters, and also cylindrical velocity
components based on proper motions, radial velocities,
and distances from \cite{24} for 72 of them.
This number also includes 45 clusters with derived
elemental abundances. We took the solar motion
relative to the local centroid to be $(U,V,W)_{\odot} =
(11.1, 12.24, 7.25)$~km/s \cite{25}, the Galactocentric
distance of the Sun to be 8.0~kpc, and the rotation
velocity of the local centroid to be 220~km/s. Table
2 (a part is presented here, while the full table
is available electronically) presents the spatial and
kinematic parameters of these clusters. The columns
of Table~2 present (1)--(2) the cluster’s name and
alternate name; (3)--(5) the coordinates ($x$, $y$, $z$) in a
right-handed Cartesian system; (6)--(7) the cluster’s
distance from the Galactic rotation axis $R_{C}$ and
Galactocentric distance $R_{GC}$; (8)--(10) the calculated
components of the space velocity $V_{R}$, $V_{\Theta}$, and $V_{Z}$ in
cylindrical coordinates, where $V_{R}$ is directed toward
the Galactic anticenter, $V_{\Theta}$ in the direction of Galactic
rotation, and $V_{Z}$ toward the North Galactic pole;
(11) the morphology index (or color) of the horizontal
branch, $HBR = (B - R)/(B + V + R)$, where B, V ,
R are the numbers of stars on the blue side of the
horizontal branch, in the instability strip, and on the
red side; (12) the absolute V magnitude; and (13) the
subsystem, where 1 denotes the thin disk, 2 the thick
disk, and 3 the halo. A letter ``G'' in column (14)
identifies clusters whose origin is associated with a
common protogalactic cloud.

In the current paper, we consider the behavior of
only four elements in globular clusters-magnesium,
silicon, calcium, and titanium-as the elements most
informative for diagnostics of early Galactic evolution.
We will pay special attention to the calcium and titanium
abundances. There are many lines of these
two elements in the visible part of the spectrum, and
their abundances can be determined fairly reliably.
The reason for our choice of these elements is that
the mean relative abundances of the two primary $\alpha$-
process elements (oxygen and magnesium) decrease
in the course of a cluster’s evolution, compared to
their abundances in the primordial protoclouds. The
abundances of another $\alpha$-process element, silicon,
were determined for a smaller number of clusters, and
are completely unknown for the field stars and dwarf
satellite galaxies that we used for comparison.

Figure~1 shows the relative abundances of all four
$\alpha$-process elements as a functioin of metallicity for
our globular clusters. For comparison, the diagrams
also display analogous relations for field stars. The
[Mg/Fe], [Ca/Fe], and [Ti/Fe] values for the stars
were taken from \cite{26}, which presents metallicities
and abundances of these elements for 785 Galactic
stars in the entire metallicity range of interest for
us. This catalog contains no [Si/Fe] ratios for these
stars, and panel (b) of Fig.~1 shows stars from the
catalog \cite{27}, which contains metallicities and relative
abundances of all $\alpha$-process elements for 714 F–G
dwarfs of the field. Unfortunately, the latter catalog
mainly contains stars belonging to the disk populations
of the Galaxy, and so has a deficiency of low metallicity
stars. In all the panels, the bars show
the dispersions of all measurements of the relative
elemental abundances in the cluster stars (note that,
to find the uncertainty of the mean for each cluster,
this dispersion must be divided by $n^{0.5}$, where $n$ is
the number of abundance estimates for the cluster
stars). We obtained the following convergences
(the dispersions of the differences between studies)
for these elements: $\langle\sigma\rm{[Si/Fe]}\rangle = 0.06$, 
$\langle\sigma\rm{[Mg/Fe]}\rangle =0.10$, 
$\langle\sigma\rm{[Ca/Fe]}\rangle = 0.08$, 
$\langle\sigma\rm{[Ti/Fe]}\rangle = 0.09$. We can
see from the panels that, in general, all four elements
in cluster stars follow those for the field stars
fairly well, indicating that an absence of systematic
errors in the abundance determinations, despite the
presence of considerable random uncertainties for the
individual clusters. The behavior of the $\alpha$-process
elements will be described in more detail below.

In order not to overload our catalog with data, we
did not include the elements of the Galactic orbits and
the cluster ages in the catalog; when necessary, the
corresponding data from \cite{28,29,30} can be used.

\section {DISTRIBUTION OF GLOBULAR
CLUSTERS OVER THE GALACTIC
SUBSYSTEMS} 

Pritzl et\,al. \cite{22} attempted for the first time to determine
the membership of globular clusters in particular
Galactic subsystems based on the components
of their residual velocities, which has long been a
practice for field stars, instead of the traditional criteria
of metallicity and horizontal-branch morphology
described above. Abundances of several elements
were found for 45 clusters in the catalog \cite{22}, for 29
of which kinematic information is available. Pritzl
et\,al. \cite{22} found that most of the clusters belong
kinematically to the Galactic halo, but a considerable
number of clusters display disk kinematics or belong
to the bulge. They assigned more than ten clusters
to the accreted halo: for these clusters, it has been
demonstrated in various studies based on their positions,
radial velocities, and, in some cases, elements
of their Galactic orbits, that they were very probably
captured from satellite galaxies, so that their origins
were intergalactic.

Clearly, there is no single, sufficient criterion for
assigning membership of globular clusters to subsystems
of the Galaxy. Reliable assignment of a
cluster to a particular subsystem requires consideration
of many parameters that are characteristic of
each subsystem, such as the cluster positions, kinematics,
metallicities, elemental abundances, ages,
and horizontal-branch morphologies. Since we are
studying chemical-composition differences for clusters
in different subsystems, we used a kinematic criterion,
using the velocity components $V_{R}$, $V_{\Theta}$, and $V_{Z}$
to calculate the probabilities of a cluster’s membership
in the thin disk, thick disk, and halo subsystems,
based on the method described in \cite{31}. This method
is similar to that applied in \cite{22}, but with somewhat
different velocity dispersions in the subsystems. Both
methods assume that the space velocity components
of stars in each of the subsystems display characteristic
normal distributions. A subsequent analysis
showed that differences in the derived memberships
for a given cluster were mainly due to differences in
the input velocities, which were more accurate in our
study. Since the subsystem membership is calculated
from residual velocities, we reduced the azimuthal
components of the cluster velocities to the orbital
velocity of the centroid at the Galactocentric distance
of the cluster. We adopted the rotation curve from
the model of the Galaxy presented in \cite{32}. Taking
into account the large distances to the clusters, resulting
in considerable uncertainties in the tangential
velocities, we calculated the probabilities of cluster
membership in the various subsystems using a recurrent
procedure. In the second step, we assumed in
the formulas used to calculate the probabilities that
the velocity dispersions and numbers of clusters in
the subsystems have the values obtained in the first
step. This resulted in a reduced inferred fraction
of objects in the thin- and thick-disk subsystems.
Though the new calculation somewhat redistributed
the memberships of a small number of clusters with
kinematics in the intermediate zones between the thin
and thick disks and between the thick disk and halo,
the overall memberships of the subsystems changed
only slightly. Changes were mainly among clusters
located near the Galactic center, where the rotation
curve is very variable. We adopted the results of this
recurrent classification procedure for our subsequent
analysis.

Figure~2a shows a Toomre diagram, 
$V_{\Theta}$ --$(V^{2}_{R}+V^{2}_{Z})^{0.5}$, 
for the globular clusters and field stars from
\cite{26}. This shows that objects displaying kinematics
of a given subsystem occupy approximately the same
area on the diagram, though the method used in \cite{26}
for the field stars differs somewhat from our method.
Using our technique, we find that the kinematic parameters
of 41 clusters result in higher probabilities
of membership in the halo than in other subsystems.
28 clusters most likely belong to
the thick disk, and thin-disk kinematics are displayed
for four clusters. (Note that, in \cite{22}, three of 29
clusters were found to diplay kinematic parameters
characteristic of the thin disk.) Among the clusters
with thick-disk kinematics, we find a considerable
number with rotation velocities around the Galactic
center even higher than the Sun's. However, the
highest azimuthal velocity is found for NGC 6553,
which displays halo kinematics: $V_{\Theta} = 383$~km/s. In
addition, more than half the halo clusters display
retrograde rotation around the Galactic center. We
believe that the origin of such clusters is extragalactic,
with high probability. According to the model of
the protogalaxy's monotonic collapse from the halo
to the disk suggested by Eggen et\,al. \cite{33}, field stars
and globular clusters genetically related to the Galaxy
cannot have retrograde orbits.

Figure~2b plots the metallicity [Fe/H] of the clusters
versus their distances from the Galactic plane $z$.
The [Fe/H] values not based on spectroscopic determinations
for other elements were taken from \cite{23},
since it contains metallicities for all the clusters. The
large circles in the figure correspond to clusters that
belong kinematically to the thin disk (light gray),
thick disk (gray), and halo (dark gray). Small filled
circles show clusters that have not been classified,
since they have unknown velocities. The most important
feature of this figure is the high concentration
of metal-rich ($\rm{[Fe/H]} > -1.0$) clusters near the
Galactic plane, independent of their membership in
the Galactic subsystems determined from kinematic
criteria. Analysis shows that the most distant points
of the orbits of all the metal-rich clusters are closer
than 5~kpc, while $Z_{max} > 10$~kpc for a considerable
fraction of he low-metallicity clusters. (An exception
is the three metal-rich clusters Pal~12, Whiting 1,
and Terzan~7, which with high probability belonged
to the disrupted Sgr dwarf galaxy earlier; see above.)
It is precisely this fact, together with the clear gap in
the metallicity at $\rm{[Fe/H]} \approx-1.0$, that suggests the
metal-rich clusters should be identified with a disk
subsystem. On the other hand, the figure shows that
half of the clusters with thin-disk kinematics (two of
the four), as well as most of the clusters with thick disk
kinematics, have $\rm{[Fe/H]} < -1.0$, at variance
with the practice of identifying disk clusters from their
high metallicities described above. It is striking that,
in the northern hemisphere ($z > 0$), three low-mass
clusters with thick-disk kinematics are located at
distances above 10~kpc from the Galactic plane. Two
of these (M 53 and Pal 5) were earlier suspected to
belong to the disrupted Sgr dwarf galaxy (see above).
There are no clusters that distant in this subsystem
in the southern hemisphere. On the other hand,
we observe a large number of distant low-metallicity
clusters with halo kinematics in the southern hemisphere,
as well as numerous clusters with unknown
velocities. This means that the observed positional
asymmetry of the kinematically selected subsystems
is not due to an observational selection effect.

The concentration of metal-rich clusters toward
the Galactic plane creates a vertical metallicity gradient,
as has been known for a long time. A plot of
distance from the Galactic plane, $|z|$, versus metallicity
(Fig.~2c) demonstrates that metal-rich clusters,
even those with halo kinematics, display a stronger
concentration toward the Galactic plane that less
metal-rich clusters with thick-disk kinematics. The
situation with the radial metallicity gradient is similar,
as is demonstrated by the plot of metallicity versus
distance from the Galactic rotation axis $R_{C}$ shown in
Fig. 2d. The mean Galactocentric distance derived
for all 47 metal-rich clusters is about 5.0~kpc, and is
a factor of three higher, 15.5~kpc, for the 110 metal poor
clusters. Note that there are no significant
correlations within each metallicity group (see \cite{18}
for details). Figure 2d shows that most of the clusters
lost by satellite galaxies are located at distances
$x \geq 5$~kpc. The space velocities of these clusters
reflect the terminal orbits of the clusters captured
from dwarf satellite galaxies disrupted by the Galaxy's
tidal forces, rather than the dynamic conditions of
star formation in the contracting protogalactic cloud.
The higher the mass of the parent satellite galaxy, the
flatter and more elongated the orbit at which it loses
its clusters and stars \cite{1}.

\section {PROPERTIES OF GLOBULAR CLUSTERS
IN DIFFERENT SUBSYSTEMS
AND WITH DIFFERENT NATURES}

Figure~3a shows a plot of the metallicity [Fe/H]
versus the azimuthal velocity $V_{\Theta}$ for globular clusters
and field stars. Different symbols correspond to different
subsystems of the Galaxy. In contrast to the similar
diagram in \cite{22}, there now appear clusters with
azimuthal velocities considerably differing from that
of the Sun in the range $\rm{[Fe/H]} > -1$. Note that all
four metal-rich clusters in retrograde orbits are within
3~kpc from the Galactic center, while the mean Galactocentric
distance for the 10 low-metallicity clusters
with $V_{\Theta} < 0$ is about 10~kpc. We also marked in the
$V_{\Theta}$--[Fe/H] diagram those clusters that were believed
at various times by various researchers to be related to
disrupted dwarf satellite galaxies (see above). We also
marked clusters located at, or having points in their
orbits ($R_{max}$) at, distances from the Galactic center
exceeding 15~kpc. We can see that there are only two
unmarked accreted clusters (NGC~2808 and $\omega$~Cen),
i.\,e., that are within this radius. An extragalactic
origin is not demonstrated for the six other clusters,
although they are located at large distances.
Noted that all the accreted and distant clusters
(with the exception of NGC~4590 and NGC~5024,
which have $V_{\Theta}$ values exceeding that of the Sun)
demonstrate a significant correlation between their
metallicities and azimuthal velocity components in
Fig.~3a (the correlation coefficient is $r = 0.66\pm0.02$).
A similar trend is present for the whole cluster sample,
though it is not statistically significant due to the considerable
scatter of the cluster azimuthal velocities for
any metallicity. The origin of this trend is the fact that
the upper velocity limit for clusters of any metallicity
is approximately constant ($V_{\Theta} \approx 350$~km/s), while
the fraction of clusters with lower velocities increases
with decreasing [Fe/H], entering the area of negative
velocities more and more. This happens expecially
abruptly at the transition across $[Fe/H] \approx -1.0$. As
a result, the velocity dispersion of the metal-rich
clusters increases abruptly, also suggesting that this
value separates the thick-disk and halo subsystems.
This relation is very significant for nearby field stars;
it is steeper and reflects Str\~omberg's asymmetric shift
due to the Galactic rotation, though our diagram plots
metallicity instead of the star's total velocity relative
to the Sun. The reason for this is that both the
total velocity and the metallicity are statistical indicators
of the ages of stellar objects in the Galaxy \cite{33}.
(Note that it is not quite correct to compare field
stars and clusters in such a diagram, due to the
fundamental difference between them: the former are
currently all located at essentially the same distance
from the Galactic center and close to the Galactic
plane, while the clusters are located at various distances.)
It may be that the increased metallicities
of the accreted clusters with higher orbital velocities
around the Galactic center follows from the fact that
the metal-richer clusters in dwarf satellite galaxies
were born closer to their centers. These clusters
thus lose their connection with their parent galaxies
when the orbits of the latter are more ``settled'' toward
the Galactic plane and their azimuthal velocities approach
the rotation velocities of the Galactic disk due
to tidal interactions with perturbations of the Galaxy's
gravitational potential, as is predicted by numerical
modeling \cite{1}. However, the presence of the correlation
in Fig.~3a and this proposed explanation must be additionally
verified because of the insufficient statistics
and considerable uncertainties in the space velocities
and Galactic orbital elements of the clusters.

Figure~3b presents a plot of the metallicity as a
function of the horizontal branch morphology HBR.
Most, but not all, of the clusters currently located
inside the solar circle mainly have extremely red or
extremely blue horizontal branches. However, some
clusters between these extreme positions are found
in a thin layer along the upper envelope in the diagram.
Note that most known accreted clusters are
located below this envelope (see the slanting line in
Fig.~3b, plotted by eye). However, this diagram
shows that this is not an absolute rule, and there are
exceptions. Distant clusters ($R_{GC}$ or $R_{max} > 15$~kpc)
and clusters in retrograde orbits ($V_{\Theta} < 0$) are also
most likely accreted. Twelve of the 22 clusters in
retrograde orbits are located inside the solar circle.
Eleven retrograde-orbit clusters have extremely blue
and three have extremely red horizontal branches; for
eight clusters, the branches are too red for their low
metallicity. As can be seen from Fig.~3b, in the range
between the extreme HBR values, all of the clusters
are below the upper envelope in the diagram. Among
the distant clusters, all except three are metal-poor,
while they can have any horizontal-branch color. It is
usually assumed that all low-metallicity clusters located
below the narrow upper band can be considered
candidate accreted clusters (see \cite{34}). This seems
very likely, suggesting we search for an explanation
for why the horizontal branches in accreted clusters
are too red.

As was noted above, recent studies show that
several episodes of star formation occurred in the
highest-mass globular clusters, with supernova outbursts
enriching the cluster's interstellar medium
with iron-peak elements. For example, several
populations with different metallicities are found in
the largest cluster, $\omega$~Centauri. However, stellar
populations with different abundances of helium
and CNO elements are also found in lower-mass
clusters (e.\,g., \cite{35}). It is supposed that the younger
star populations in these latter clusters were formed
from chemically contaminated matter ejected by
intermediate-mass giants on the AGB, rapidly rotating
massive stars, and rotating first-generation
AGB stars \cite{36,37}. In the course of time, extended
horizontal branches are formed in such clusters, and
as a result, their color no longer corresponds to the
primordial low-metallicity chemical composition of
the stellar population that dominates by number. The
numerical simulations of Jang et\,al. \cite{38} demonstrated
that, indeed, the horizontal-branch color
becomes redder in clusters with secondary younger
populations, enriched mainly with CNO elements.
The cluster's Oosterhoff type changes simultaneously.
According to models, all this happens at early
stages of the cluster's evolution, within one billion
years after the last star-formation burst.

Figure~3b shows that, among the 41 clusters with
extremely blue horizontal branches ($\rm{HBR} > 0.85$),
only eight are located or have orbital points at distances
more than 15~kpc from the Galactic center.
At the same time, 29 clusters are currently inside the
solar circle ($R_{GC} < 8$~kpc), and four clusters are between
these boundaries. Note that almost all the distant
clusters are fairly faint in terms of their absolute
magnitudes, i.\,e., they have low masses. This is clearly
demonstrated by Fig.~3c, where we display the absolute
magnitude $M_{V}$ as a function of distance from
the Galactic center $R_{GC}$ for clusters with extremely
blue horizontal branches. All the clusters currently
located at distances from the Galactic center exceeding
15~kpc have $M_{V} \geq -8.0^{m}$ (apart from the distant,
bright cluster NGC~2419 with the boundary value,
HBR = 0.86), while all the brighter clusters are relatively
nearby. (Note, however, that it has long been
known that low-mass clusters dominate among distant
clusters (see \cite{39} and references therein), while
this rule is expressed more clearly for blue clusters.)
The result is that extremely blue horizontal branches
are observed mainly among low-metallicity clusters
that are close to the Galactic center, as well as for a
small number of distant clusters with comparatively
low masses. We believe that this could be due to the
fact that, for both types of clusters, matter ejected
by evolved stars does not stay in the clusters, and is
swept away by perturbations of the Galaxy's gravitation
potential. In the case of low-metallicity clusters
that are close to the Galactic center, this occurs due
to frequent approaches toward the Galaxy's bulge
and disk; distant clusters with comparatively low
masses are unable to retain this ejected matter even
at considerable distances from the Galactic center,
due to their low masses. As a result, they do not
form secondary populations, or form them with only
a small number of stars. Low-mass clusters with
reddened horizontal branches often have all points of
their orbits outside the solar circle, where the effects
from the Galaxy's gravitation potential are reduced.
This may be why they have time to form a population
of younger stars that distort the colors of their horizontal
branches. The picture we have described is
not entirely straightforward, because the third and
subsequent populations in some clusters are over-enriched
in helium, resulting in the horizontal-branch
stars appearing toward the high-temperature side of
the instability strip. As a consequence, the color of
the horizontal branch is displaced toward the blue.
A typical example is M 15 where, in addition to the
normal blue part of the horizontal branch, there also
is a so-called ``blue tail'' \cite{40}. Testing our proposed
explanation for the existence of a correlation between
the color of the horizontal branch and the loss of gas
by the cluster requires a detailed analysis of cluster
orbital tracks, as well as taking into account numerous
previously published data on the chemical
compositions of individual cluster stars.

Figure 3d plots the [Ca, Ti/Fe] ratios versus the
azimuthal velocity $V_{\Theta}$ for globular clusters of our
sample for which these parameters available and for
field stars. Clusters demonstrated to have an extragalactic
origin, i.e. those believed to be accreted,
are marked, as well as clusters with orbital points
($R_{max}$) or positions at distances from the Galactic
center exceeding 15~kpc. The vertical line at $V_{\Theta} =0$ 
separates field stars and clusters with retrograde
rotation. On average, field stars typically have high
[$\alpha$/Fe] ratios, but with a large scatter for low and
negative velocities, which rapidly decrease with approach
to the rotation velocity of the Galactic disk at
the solar Galactocentric distance. The [$\alpha$/Fe] ratios
for globular clusters displaying any type of kinematics
do not differ strongly, and show absolutely no correlation
with the azimuthal velocity component, unlike
the metallicity in Fig.~3a. For all $V_{\Theta} < V_{\odot}$, their
dispersion is not large 
($\sigma\rm{[\alpha/Fe]} \approx 0.1$); however, the
scatter increases greatly for orbital velocities around
the Galactic center exceeding the solar value (there
are only five such clusters, but with clusters displaying
kinematics of all three subsystems among them).
The high relative abundances of $\alpha$-process elements
indicate that almost all these clusters were formed
from interstellar matter that was not yet enriched
with iron-peak elements due to Type Ia supernova
outbursts.

\section {RELATIVE ABUNDANCES
OF $\alpha$-PROCESS ELEMENTS IN GLOBULAR
CLUSTERS OF DIFFERENT SUBSYSTEMS
AND WITH DIFFERENT NATURES}

Figure~4a presents plots of [Ca,Ti/Fe] versus
[Fe/H] for globular clusters of different Galactic
subsystems and for field stars with different natures
(details will be described below). This figure shows
that, in contrast to the field stars, clusters belonging
kinematically to each of the subsystems can have
any metallicity, and also any relative abundances
of $\alpha$-process elements. We can see from Fig.~4b,
which shows such a diagram plotted for the averaged
abundances of four $\alpha$-process elements (magnesium,
silicon, calcium, and titanium), that, generally speaking,
the position of the area occupied by the globular
clusters does not change with respect to the field
stars, but the number of clusters has decreased.
In contrast to the other panels of the figure, this
panel uses different symbols for field stars of different
Galactic subsystems identified using the kinematic
criterion of \cite{31}. The clusters and field stars from a
given subsystem have considerably different chemical
compositions. For genetically related field stars,
i.e. those formed from the general protogalactic
cloud, metallicity can be a statistical indicator of their
ages, since the general abundance of heavy elements
steadily increases with time in a confined star and
gas system (such as, in a first approximation, our
Galaxy). We believe that these are the field stars
with residual velocities $V_{res} < 240$~km/s (see \cite{41}),
plotted in the diagram as small dark asterisks. The
vast majority of field stars with higher residual velocities
(gray pluses) show retrograde rotation (see
Fig.~3a). All the stars with higher velocities can be
considered candidate accreted stars. Note that the
low-metallicity ($\rm{[Fe/H]} < -1.0$) genetically related
field stars have positions along the top half of the
strip in Figs. 4a, 4c, and 4d. For orientation in the
figure, we plotted by eye a broken line corresponding
to the lower envelope for the genetically related field
stars. The position of our line is in good agreement
with that plotted in \cite{42}, where two populations were
identified among the low-metallicity field stars not
from their kinematics, but from their relative abundances
of $\alpha$-process elements, with the boundary
approximately at $[\alpha/Fe] \sim 0.3$, and it was found that
these populations of stars differ not only in their
chemical compositions but also their kinematics and
ages. Note that Shetrone et\,al. \cite{42} were originally
looking for evidence that a population with lower
relative abundances of $\alpha$-process elements had an
extragalactic origin. Figures 4a, c, and d show
that the [$\alpha$/Fe] ratios for genetically related stars
begin to abruptly decrease with increasing metallicity
starting from $\rm{[Fe/H]} \approx -1.0$, due to the onset of
Type Ia supernova outbursts in the Galaxy. This is not
observed for globular clusters, and the vast majority
of metal-rich clusters are above the strip occupied
by field stars. Though some decrease of the [$\alpha$/Fe]
ratios with increasing metallicity can be noted for
them, their positions in the diagram mainly remain
in the range $\rm{[\alpha/Fe]} > 0.15$, as for the low-metallicity
clusters. Note that clusters kinematically belonging
to the two most populous Galactic subsystems, the
thick disk and halo, exhibit no statistically significant
differences in their positions in the diagram.

Figure~4c displays the same [Ca, Ti/Fe] versus
[Fe/H] plot but with the clusters identified using other
criteria: accreted clusters whose relationship to satellite
galaxies disrupted in the past has been established
from their positions and space motions, distant
clusters ($R_{GC}$ or $R_{max} > 15$~kpc), and clusters in
retrograde orbits. The clusters in the range $\rm{[Fe/H]} <
-1.0$ are located in the diagram such that the lower
envelope for the genetically related field stars is close
to their median. In general, the entire population of
accreted clusters, together with candidate accreted
clusters (distant clusters and clusters in retrograde
orbits), exhibit a large scatter in their [$\alpha$/Fe] ratios
in Fig.~4c. (Strikingly, five of the nine clusters in
retrograde orbits were found inside the solar circle.)
The scatter for these clusters is considerably higher
than for the genetically related field stars. Approximately
the same strong scatter is exhibited by the
high-velocity ($V_{res} > 240$~km/s), low-metallicity field
stars in Fig. 4, which are not genetically related to
the general protogalactic cloud and probably have an
extragalactic origin. It may be that the large scatter in
the [$\alpha$/Fe] ratios for such clusters and the field stars
could arise due to different maximum masses for the
Type II supernovae that have enriched matter in their
numerous parent dwarf galaxies.

The black circles in Fig.~4d show clusters that
cannot be considered to be candidate accreted clusters
by any criteria. We assume such clusters to
be genetically related, i.\,e. formed from the general
protogalactic cloud. By definition, all 32 such clusters
in our sample are closer than 15~kpc from the
Galactic center, and 27 of them, plotted as white
triangles inside dark circles, are even inside the solar
circle ($R_{GC} < 8$~kpc). Among them, we find all rich
clusters with high [Ca,Ti/Fe] ratios, some of which
probably belong to the Galactic bulge. In addition
to the genetically related clusters, this figure shows
clusters with relationships to two fairly high-mass
dwarf galaxies, Sgr and CMa, that are believed to be
reliably established (see above). We can see that 24 of
the 28 accreted and genetically related clusters in the
range $\rm{[Fe/H]} < -1.0$ form a fairly narrow strip in the
diagram, so that the lower envelope for the genetically
related field stars could also be a lower envelope for
them. However, all these clusters are more closely
concentrated toward the line than the genetically related
field stars. (Very low [Ca, Ti/Fe] ratios are
displayed by only two low-mass clusters from the
Sgr galaxy: the very distant cluster NGC~2419 with
only one studied star and the cluster Ter~8. However,
the relative magnesium abundances are high
for both of these, 0.30 and 0.52, respectively, and
the silicon abundance for Ter~8 is 0.38; i.\,e., taking
into account all the $\alpha$-process elements, these two
clusters also appear near the lower envelope.) On the
other hand, in the metal-richer range, both metal-rich
clusters captured from the Sgr dwarf satellite
galaxy (Pal~12 and Ter~7) lie below the field stars.
The low-metallicity cluster Rup~106, believed to be
lost by a dwarf galaxy, also has an abnormally low
position in the diagram. This cluster was proposed
with some probability in \cite{12} to belong to the CMa
galaxy, which was fairly massive in the past, but its
very low relative abundance of $\alpha$-process elements
and low metallicity contradict this hypothesis. It may
be that it was lost instead by one of low-mass satellite
galaxies, if the abundances of $\alpha$-process elements for
only two stars in Rup~106 published in one paper
are correct. Note that this is one of the lowest-mass
($M_{V} = -6.35^{m}$) low-metallicity clusters in the
Galaxy, and could plausibly have originated in such a
dwarf galaxy.

\section {ACCRETED GLOBULAR CLUSTERS
AND MASSES OF THEIR PARENT
GALAXIES} 

A relation between [Mg, Ca/Fe] and [Fe/H] was
derived in \cite{46} for 235~stars selected in that paper
as origination in the core of the currently disrupting
Sgr dwarf galaxy. It is emphasized that the sequence
for stars from this galaxy in the low-metallicity range
($\rm{[Fe/H]} < -1.0$) coincides with that for Galactic field
stars, while it is somewhat lower than field stars with
higher metallicities. Mucciarelli et\,al. [46] remark
that, in the range $\rm{[Fe/H]} > -1.0$, the metallicity relation
for the relative $\alpha$-process element abundances in
the Sgr galaxy is very similar to that observed for stars
in the highest-mass satellite of our Galaxy, the Large
Magellanic Cloud. In their opinion, this suggests a
high mass also for the Sgr galaxy. Indeed, it was
demonstrated in \cite{47} from modeling of the kinematics
of the stellar tidal tail of the Sgr galaxy that the mass
of its dark halo should be $M = 6 \ast 10^{10} M_{\odot}$ in order
to reproduce the velocity dispersion in this galaxy's
stream. Mucciarelli et\,al. \cite{46} were able to reproduce
the observed chemical properties of the parent
Sgr dwarf galaxy in a model assuming a comparably
high initial mass and a considerable loss of mass
several billion years ago, in the period starting with
the galaxy's first crossing of our Galaxy's pericenter.

The large gray crosses in Fig. 4d identify stars
of the so-called Centaurus stream among the field
stars. It is supposed that all these stars were lost
by the dwarf satellite galaxy whose central core was
the highest-mass globular cluster $\omega$~Centauri, which
now belongs to our Galaxy (see \cite{48} and references
therein). Numerical modeling of dynamical processes
during the interaction of the satellite galaxy with our
Galaxy's disk and bulge demonstrates that capturing
the core of a dwarf galaxy onto an eccentric retrograde
orbit with a low apogalactic radius is quite possible,
provided that the galaxy had a fairly high mass, 
$\approx10^{9} M_{\bigodot}$ \cite{14}. In particular, the numerical modeling
of Abadi et\,al. \cite{49} demonstrates that the sizes of the
orbits of sufficiently massive satellite galaxies steadily
decrease, and are moved toward the Galactic plane by
dynamical friction. With time, such galaxies obtain
very eccentric orbits parallel to the Galactic disk, and
the Galaxy's tidal forces begin to effectively disrupt
them during each of their passages through the perigalactic
distance, so that they lose stars with a certain
well defined orbital energy and angular momentum.
Thus, if the observer's position is between the
apogalactic and perigalactic radii of such an orbit, the
tidal ``tail'' of the disrupting galaxy will be observed as
a ``moving group'' of stars with small vertical velocity
components and a broad, symmetric, and often two-peaked
distribution of the radial components of the
space velocities.

Based on the recommendations of \cite{50}, Marsakov
and Borkova \cite{41} identified from their original catalog
of spectroscopically determined Mg abundances
(an $\alpha$-process element) in $\approx 800$ nearby F--K field
dwarfs \cite{51} stars lost by the dwarf galaxy whose core
had been $\omega$~Cen, with the azimuthal and vertical components
of their velocities in the ranges $-50 \leq V_{\Theta} \leq0$~km/s 
and $|V_{Z}| < 65$~km/s. The identified 18 stars
of the stream were found to form a fairly narrow
sequence in the [Fe/H]--[Mg/Fe] diagram, characteristic
of genetically related stars. The position of the
``break point'' of the relative magnesium abundance at
$\rm{[Fe/H]} \approx -1.3$ dex indicates that the star-formation
rate was lower in the parent galaxy than in our Galaxy.
The star formation in this galaxy apparently lasted
so long that its metal-richest stars had reached the
ratio $\rm{[Mg/Fe]} < 0.0$ dex, even lower than the solar
value. However, the low maximum metallicity for
the stars in this group (only $\rm{[Fe/H]} \approx -0.7$) 
indicates that subsequent star formation in their parent
galaxy had ended. This probably happened because
the dwarf galaxy began to be disrupted. In other
words, the chemical composition of the stars in this
former galaxy indicates that it evolved over a long
time (although shorter than our Galaxy) before being
disrupted. We applied the same criteria to identify
stars of the Centaurus stream in the catalog of field
stars used in the current study \cite{26}. There are 18
such stars, plotted as large crosses in Fig.~4d. The
behavior of two other $\alpha$-process elements (calcium
and titanium) corresponds to the behavior of magnesium
according to data from another catalog. As
a result, we find that the relationship between [$\alpha$/Fe]
and [Fe/H] for stars of the Centaurus stream agrees
reasonably well with that for the accreted clusters in
the range $\rm{[Fe/H]} > -1.5$. Thus, the hypothesis of
an intergalactic origin is confirmed, at least for some
of the high-velocity field stars that came to us from
satellite galaxies with fairly high masses.

\section {CONCLUSIONS} 

Our Galaxy possesses a complex, multicomponent
structure consisting of several subsystems that
are, in some sense, embedded in each other. There
are no clear boundaries between the subsystems, and
their sizes can be estimated only approximately. Inferred
geometric boundaries assume certain velocity
dispersions for objects belonging to a given subsystem.
It is believed that using kinematic parameters
is the most reliable method for distributing objects
among the subsystems. This particular method
was used in order to distribute field stars among the
Galaxy's subsystems. The results of our analysis
show that this method is poorly applicable to globular
clusters, because clusters of different subsystems
identified kinematically demonstrate properties
of their chemical compositions that are fundamentally
different from those of field stars in the same
subsystem, and vice versa. In particular, all metal-rich
($\rm{[Fe/H]} > -1.0$) clusters belonging kinematically
to any of the subsystems are confined within
fairly restricted limits about the Galactic center and
Galactic plane. At the same time, there are fairly
distant metal-poorer clusters among all kinematically
identified subsystems. This is manifest through
the well-known radial and vertical metallicity gradients
in the globular-cluster population of the Galaxy.
Thus, we find the traditional procedure of distinguishing
thick-disk clusters from halo clusters according
to their metallicity to be more acceptable.
(Note that a similar discrepancy between criteria for
membership in the thick-disk and halo subsystems
based on chemical and kinematic properties is also
observed for field RR Lyrae stars; see, in particular,
\cite{52}.) Recall that the probabilities of cluster
membership in the Galactic subsystems were calculated
from the clusters' residual velocities at the
Galactocentric distances corresponding to their current
positions. However, we did not take into account
how high above the Galactic plane the clusters are
now. As a result, the vertical components of the
residual velocities of clusters that are far from the
Galactic plane may be underestimated, since these
components become lower near the apogalactic point.
In turn, this could result in an erroneous classification
of such clusters as objects of the disk subsystem. We
are planning to consider this circumstance in a future
paper and to perform a refined classification of all the
clusters.

If all globular clusters formed from matter of a
single protogalactic cloud, we must suppose that
it is the existence of active phases in the Galaxy's
evolution that is responsible for the special position
of metal-rich clusters (see \cite{19}). This active phase
would begin after a large number of supernova outbursts
in the halo that heat the interstellar matter,
resulting in a delay of star formation. During this
delay, the protogalaxy's interstellar matter, already
contaminated with heavy elements, mixes, cools, and
collapses to a smaller size, after which the disk subsystems
of the Galaxy are formed. However, as is
shown in Figs.~4a--4d, this scenario for the formation
of globular-cluster subsystems is in a contradiction
with the relative abundances of $\alpha$-process elements
in the clusters, which were found to be high for
almost all the studied metal-rich clusters (with the
exception of the three accreted clusters Ter~7, Pal~12,
and Rup~106 and the two bulge clusters NGC~6528
and NGC~6553): $\rm{[\alpha/Fe]} > 0.15$. The absence of a
well-defined ``bend'' in the relation between [$\alpha$/Fe]
and [Fe/H], as is present for field stars, indicates
that all the studied clusters were formed before the
onset of Type Ia supernova outbursts, during the first
billion years after the beginning of star formation in
the protogalactic cloud. These supernovae enrich the
interstellar medium exclusively with atoms of iron-peak
elements; as a result, the [$\alpha$/Fe] ratios in the
closed star and gas system begin to decrease. As
demonstrated by the field stars in Figs. 4a--4d, this
happens in the Galaxy at $\rm{[Fe/H]} \approx -1.0$. This figure
also shows that, within the metal-rich range, the
clusters also exhibit a decrease of the relative abundances
of $\alpha$-process elements with increasing metallicity,
but, at any metallicity, their [$\alpha$/Fe] ratios remain
higher than those for thick-disk field stars. The result
is that their relation between [$\alpha$/Fe] and [Fe/H] lies
parallel to and above the analogous relation for the
field stars. Note that clusters of all kinematically
identified subsystems are present among them. Figure
4d shows that all metal-rich clusters are inside
the solar circle. Even the most distant points of their
orbits just barely cross this Galactocentric radius.
Because of their uncertainty, inferred cluster ages do
not admit definite conclusions about their natures.
In particular, according to the ages of \cite{29}, all are
younger than 12~billion years. However, the estimates
of \cite{30} suggest that metal-rich globular clusters are
older than this, and were formed simultaneously with
the oldest, lowest-metallicity clusters. Thus, there is
no consistent explanation of why the clusters abruptly
change the volume they occupy in the Galaxy when
crossing the metallicity boundary $\rm{[Fe/H]} \approx -1.0$.

Our Fig.\,4 shows that the whole sample of the
Galaxy's low-metallicity ($\rm{[Fe/H]} < -1.0$) globular
clusters occupies essentially the same strip in the
[Fe/H]--[$\alpha$/Fe] diagram as the high-velocity ($V_{\Theta} >
240$~km/s), i.e. accreted, field stars. We can see from
this same diagram that stars of dwarf galaxies that
are satellites of the Galaxy have much lower [$\alpha$/Fe]
ratios at the same low metallicity \cite{42,43,44}. This
indicates that all stellar objects of the accreted halo
are remnants of galaxies with higher masses than the
present environment of the Galaxy. The difference in
$\alpha$-process element abundances for Galactic stars and
stars in lower-mass dwarf satellite galaxies testifies
that the latter stars have not left appreciable traces
in the Galaxy. This agrees with the conclusions
drawn in \cite{22} based on a smaller number of globular
clusters. The recent paper \cite{53} also concludes that a
high-mass satellite was accreted by the Galaxy some
(8--11) billion years ago, based on the detection of a
strong radial anisotropy of the velocity field for a large
sample of halo dwarfs within $\sim 10$~kpc of the Sun.

%%%%%%ACKNOWLEDGMENTS%%%%%%%%%%
\section*{ACKNOWLEDGMENTS}

This work was supported by the Ministry of
Science and Education of the Russian Federation
(State Contracts No. 3.5602.2017/BCh and
No. 3.858.2017/4.6).

\renewcommand{\refname}{REFERENCES}

\newpage

\begin{figure*}
\centering
\includegraphics[angle=0,width=0.99\textwidth,clip]{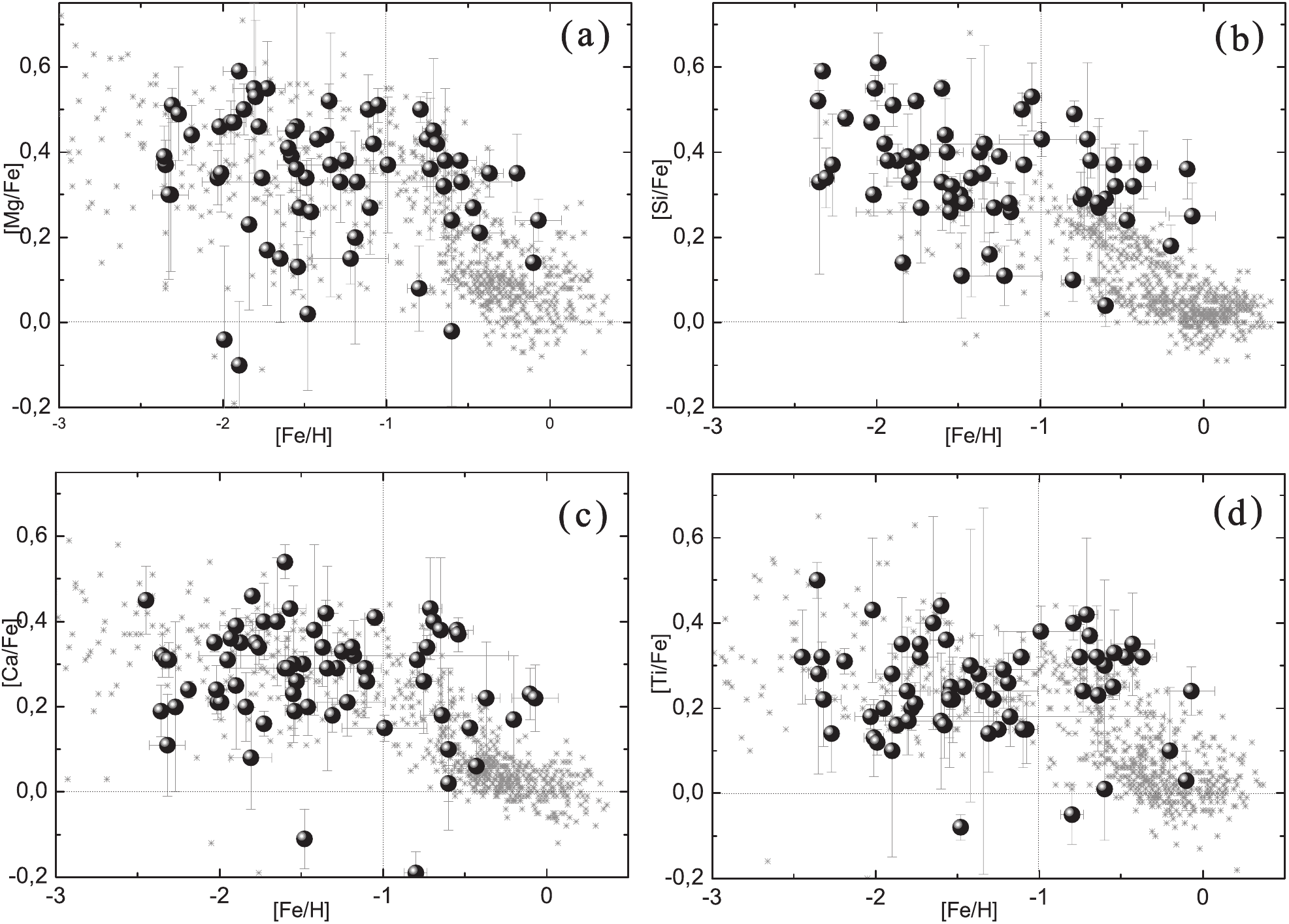}
\caption{Metallicity dependences for the relative abundances 
         of (a) magnesium, (b) silicon, (c) calcium, and (d) 
         titanium for globular clusters from our catalog and 
         for field stars from \cite{26} [(a), (c), (d)] 
         and \cite{27} (b). The filled circles are globular 
         clusters and the gray asterisks are field stars.}
\label{fig1}
\end{figure*}

\newpage

\begin{figure*}
\centering
\includegraphics[angle=0,width=0.99\textwidth,clip]{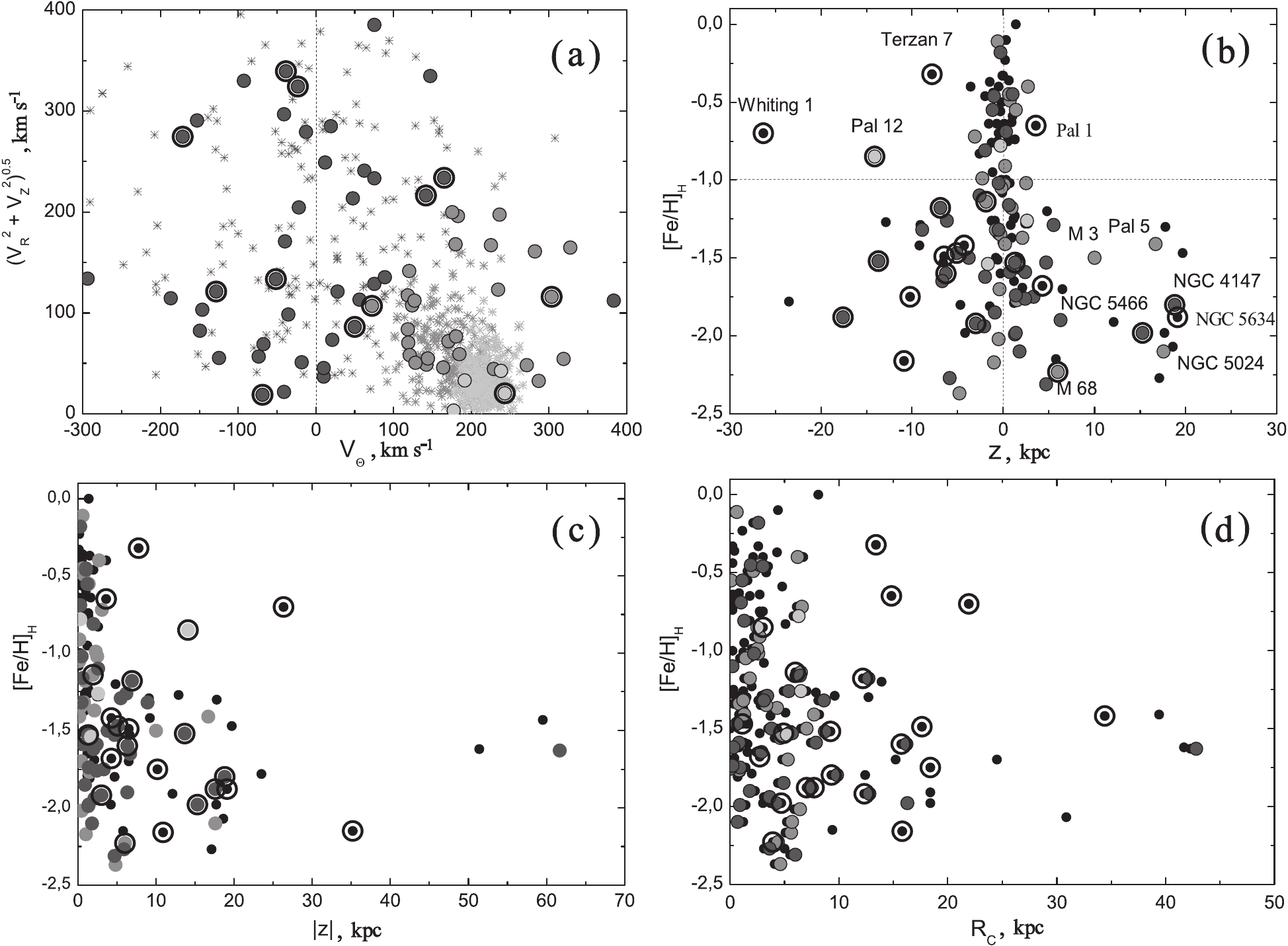}
\caption{(a) Toomre diagram for globular clusters and 
         field stars from \cite{26}. Metallicity versus (b) 
         distance of the globular clusters from the Galactic 
         plane, (c) absolute value of the distance from the 
         Galactic plane, and (d) distance from the Galactic 
         rotation axis. Field stars located in the thin 
         disk are plotted as faint gray asterisks, those in 
         the thick disk as gray crosses, and those in the 
         halo as black asterisks. Globular clusters are 
         plotted as filled circles of the corresponding 
         colors for the three subsystems. Circled clusters 
         are those known to have been lost by dwarf galaxies. 
         The small filled circles are clusters not attributed to 
         subsystems. The [Fe/H] values were taken 
         from the catalog \cite{23}.}
\label{fig2}
\end{figure*}

\newpage

\begin{figure*}
\centering
\includegraphics[angle=0,width=0.99\textwidth,clip]{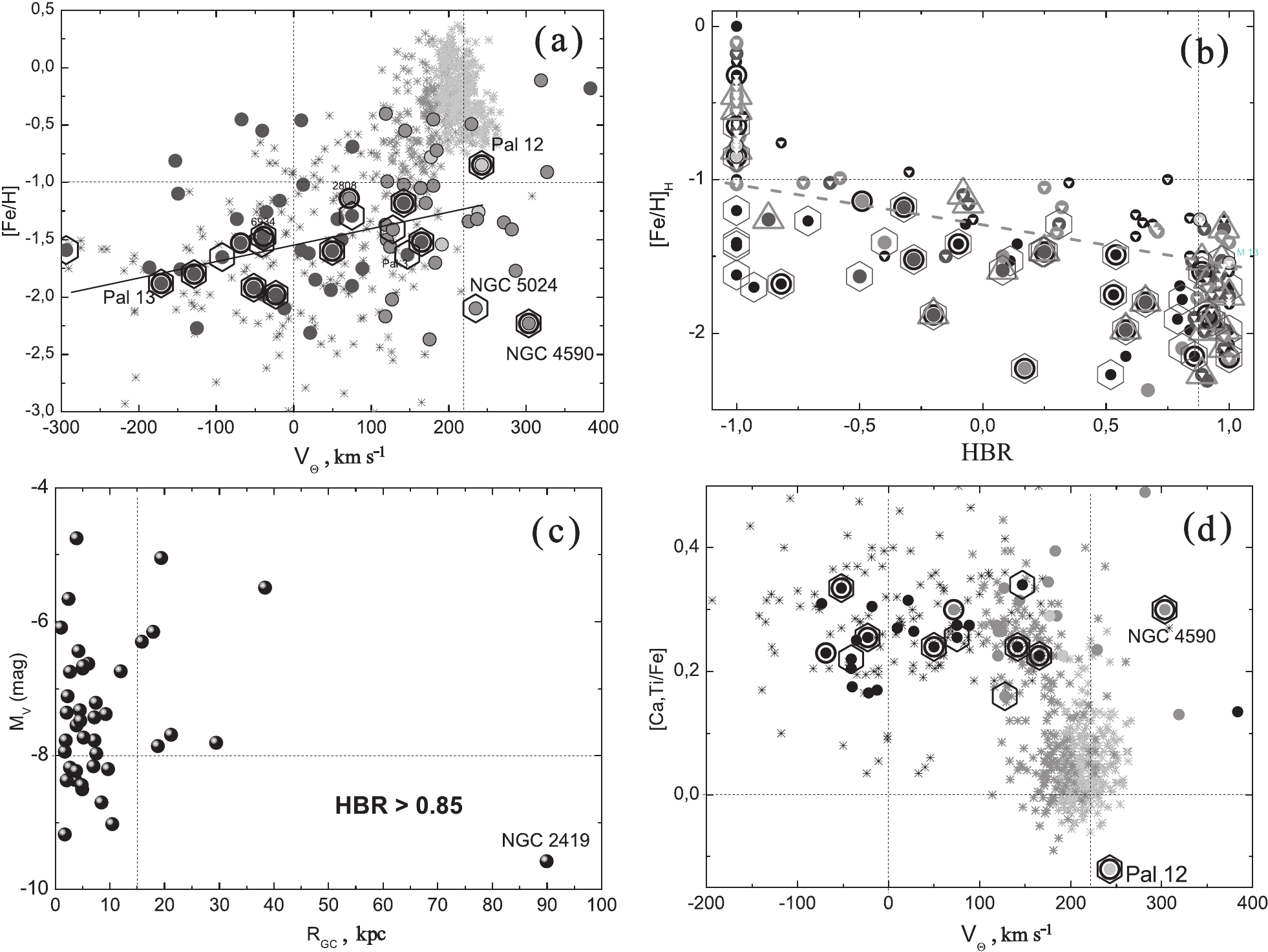}
\caption{(a) Spectroscopicmetallicity versus the rotation 
         velocity around the Galactic center for field stars 
         and globular clusters; (b) metallicity from the 
         catalog \cite{23} versus the color of the horizontal 
         branch; (c) absolute magnitude versus Galactocentric
         distance for clusters with extremely blue horizontal 
         branches; (d) relative abundances of $\alpha$-process 
         elements versus the cluster rotation velocities 
         around the Galactic center. In panels (a), (b), and 
         (d), large, filled hexagons around circles denote 
         distant clusters ($R_{G}$ or $R_{max} > 15$~kpc), 
         light gray triangles around circles clusters in 
         retrograde orbits, and white triangles inside circles 
         clusters located inside the solar circle ($R_{G} < 8$~kpc). 
         The dashed horizontal lines are at [Fe/H] = –1.0 (a), 
         (b) and [$\alpha$/Fe] = 0.0 (d); the dotted vertical 
         lines in (a) and (d) correspond to $V_{\Theta} = 0$ 
         and 220~km/s and those in (b) to HBR = 0.85; the 
         slanting dashed line was drawn by eye and separates 
         the positions of the inner and outer clusters. The 
         remaining symbols are as in Fig.~2. The names of 
         clusters strongly deviating from the mean positions 
         for the corresponding subsystems are indicated.}
\label{fig3}
\end{figure*}

\newpage

\begin{figure*}
\centering
\includegraphics[angle=0,width=0.99\textwidth,clip]{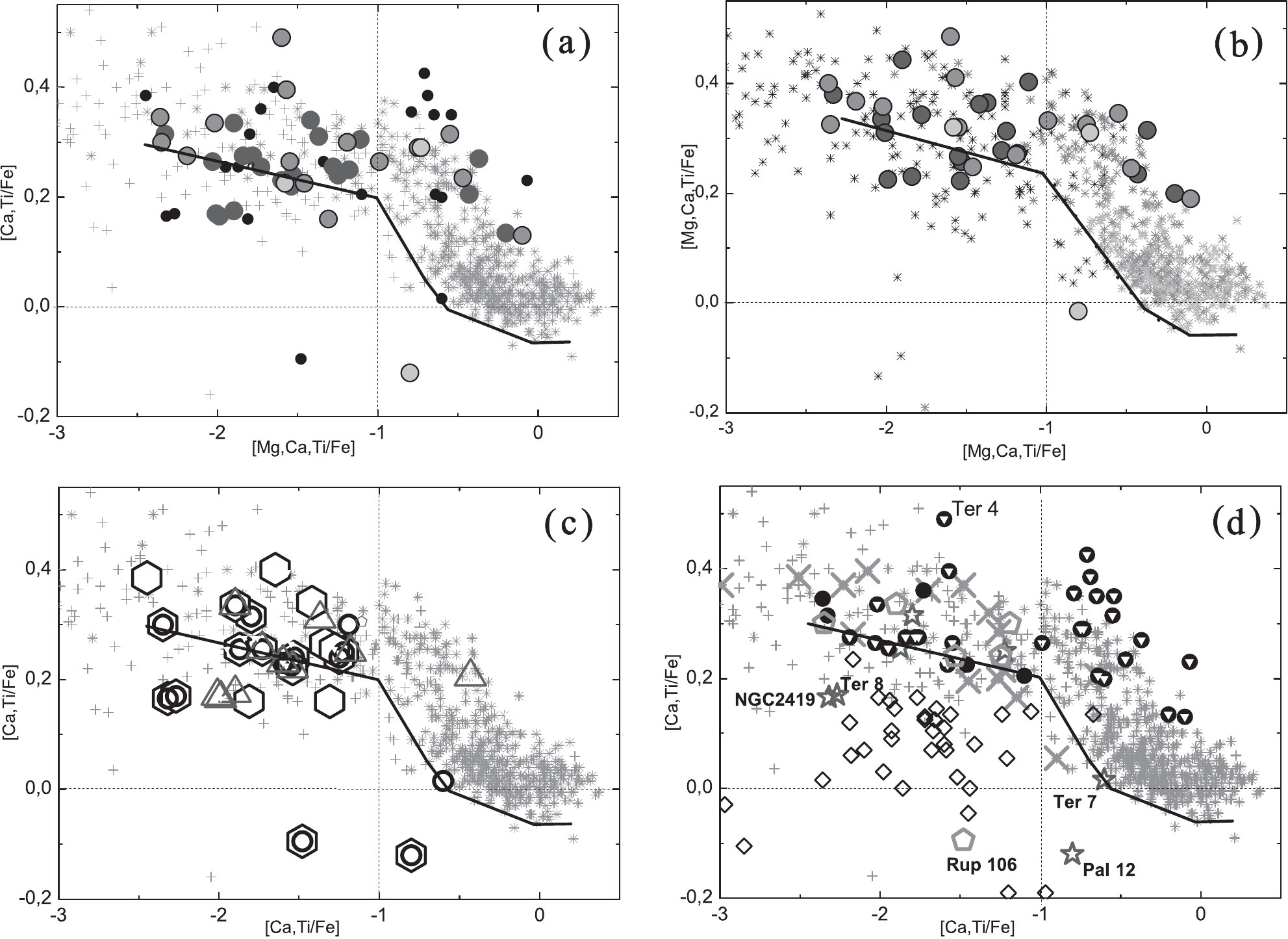}
\caption{Relative abundances averaged over [(a), (c), (d)] 
         two $\alpha$-process elements (Ca and Ti) and (b) 
         four $\alpha$-process elements (Mg, Ca, Si, and Ti) 
         (b) versus metallicity for field stars from \cite{26} 
         and for globular clusters (since Si abundances 
         are not available for the field stars, their 
         abubdances were averaged over the remaining three 
         elements). The symbols for field stars and clusters 
         in different subsystems are the same as in Figs.~2a,\,2b. 
         Dark asterisks show genetically related field stars with
         $V_{res} < 240$~km/s, and faint gray pluses field 
         stars with higher velocities (c), (d). The symbols 
         for outer, inner, and retrograde orbit clusters are 
         as in Fig. 3c. The large gray pentagons show clusters 
         lost by the CMa galaxy, the large gray stars clusters
         lost by the Sgr galaxy, diamonds stars from the 
         dwarf satellite galaxies of \cite{42,43,44}, and 
         large crosses stars of the Centaurus stream (d). 
         The broken line, drawn by eye, is the lower envelope 
         for the genetically related field stars (a)–(d).}
\label{fig4}
\end{figure*}

\newpage
\clearpage

\newpage
\begin{landscape}

\begin{table}[t!]

\caption{%
Mean relative elemental abundances and the abundance dispersions 
($\sigma$) derived from stars in globular clusters 
(Full version of the catalog:  
http://cdsarc.u-strasbg.fr/viz-bin/qcat?J/AZh/96/267)}
\bigskip
%\label{tproto}
\begin{center}
\begin{tabular}{c|c|c|c|c|c|c|c|c|c|c|c|c|c|c|c}
\hline \hline

\multicolumn{1}{c|}{\parbox{1.6cm}{Name}}&
\multicolumn{1}{c|}{\parbox{1.0cm}{Other name}}&
\multicolumn{1}{c|}{\parbox{1.0cm}{[Fe/H]}}&
\multicolumn{1}{c|}{\parbox{0.7cm}{$\sigma$}}&
\multicolumn{1}{c|}{\parbox{1.0cm}{[C/Fe]}}&
\multicolumn{1}{c|}{\parbox{0.3cm}{$\sigma$}}&
\multicolumn{1}{c|}{\parbox{1.0cm}{[O/Fe]}}&
\multicolumn{1}{c|}{\parbox{0.8cm}{$\sigma$}}&
\multicolumn{1}{c|}{\parbox{1.0cm}{[Na/Fe]}}&
\multicolumn{1}{c|}{\parbox{0.8cm}{$\sigma$}}&
\multicolumn{1}{c|}{\parbox{1.0cm}{[Mg/Fe]}}&
\multicolumn{1}{c|}{\parbox{0.8cm}{$\sigma$}}&
\multicolumn{1}{c|}{\parbox{1.0cm}{[Al/Fe]}}&
\multicolumn{1}{c|}{\parbox{0.8cm}{$\sigma$}}&
\multicolumn{1}{c|}{\parbox{1.0cm}{[Si/Fe]}}&
\multicolumn{1}{c}{\parbox{0.8cm}{$\sigma$}}\\

\hline
   1    &  2   & 3   &  4 & 5& 6& 7  &  8 & 9  & 10 & 11 & 12 & 13 & 14 &15&16\\
\hline
NGC 104 &47 Tuc&-0.75&0.08& -& -&0.18&0.08&0.33&0.18&0.43&0.11&0.37&0.21&0.29&0.06\\
NGC 288 &      &-1.37&0.04& -& -&0.34&0.14&0.24&0.06&0.44&0.01&0.45&0.06&0.40&0.04\\
NGC 362 &      & -1.2&0.08& -& -&0.76&0.53&0.11&0.07&0.33&0.02&0.27&0.05&0.26&0.07\\
NGC 1261&      &    -&   -& -& -&   -&   -&   -&   -&   -&   -&   -&   -&-&-\\
NGC 1851&      &-1.25&0.07& -& -&0.03&0.21&0.20&0.04&0.38&0.02&0.38&0.20&0.39&0.01\\
\hline

\end{tabular}
\end{center}

%\label{tproto}
\begin{center}
\begin{tabular}{c|c|c|c|c|c|c|c|c|c|c|c|c|c|c|c|c}

\hline \hline

\multicolumn{1}{c|}{\parbox{1.6cm}{Name}}&
\multicolumn{1}{c|}{\parbox{1.0cm}{[K/Fe]}}&
\multicolumn{1}{c|}{\parbox{0.4cm}{$\sigma$}}&
\multicolumn{1}{c|}{\parbox{1.0cm}{[Ca/Fe]}}&
\multicolumn{1}{c|}{\parbox{0.8cm}{$\sigma$}}&
\multicolumn{1}{c|}{\parbox{1.0cm}{[Sc/Fe]}}&
\multicolumn{1}{c|}{\parbox{0.8cm}{$\sigma$}}&
\multicolumn{1}{c|}{\parbox{1.0cm}{[Ti/Fe]}}&
\multicolumn{1}{c|}{\parbox{0.8cm}{$\sigma$}}&
\multicolumn{1}{c|}{\parbox{1.0cm}{[V/Fe]}}&
\multicolumn{1}{c|}{\parbox{0.8cm}{$\sigma$}}&
\multicolumn{1}{c|}{\parbox{1.0cm}{[Cr/Fe]}}&
\multicolumn{1}{c|}{\parbox{0.8cm}{$\sigma$}}&
\multicolumn{1}{c|}{\parbox{1.0cm}{[Mn/Fe]}}&
\multicolumn{1}{c|}{\parbox{0.8cm}{$\sigma$}}&
\multicolumn{1}{c|}{\parbox{1.0cm}{[Co/Fe]}}&
\multicolumn{1}{c}{\parbox{0.8cm}{$\sigma$}}\\

\hline
   1    & 17 &18& 19 & 20 & 21  & 22 & 23 & 24 & 25  & 26 & 27 & 28 & 29&30&31&32\\

\hline
NGC 104 &-0.6&-&0.26&0.12& 0.15&0.11&0.32&0.08& 0.12&0.08&-0.03&   -&-0.24&-0.06&0.00&-\\
NGC 288 & -  &-&0.34&0.09& 0.02&0.12&0.28&0.09& 0.00&0.05&    -&   -&   -&-&-&-\\
NGC 362 & -  &-&0.32&0.08&-0.07&0.03&0.18&0.07&-0.05&0.01&-0.03&0.01&-0.33&-&-0.11&0.16\\
NGC 1261& -  &-&   -&   -&    -&   -&   -&   -&    -&   -&    -&   -&   -&-&-&-\\
NGC 1851& -  &-&0.33&0.02& 0.03&0.03&0.15&0.01&-0.12&0.05&-0.02&0.06&-0.34&0.08&-0.07&0.15\\
\hline

\end{tabular}
\end{center}

%\label{tproto}
\begin{center}
\begin{tabular}{c|c|c|c|c|c|c|c|c|c|c|c|c|c|c|c|c}

\hline \hline

\multicolumn{1}{c|}{\parbox{1.6cm}{Name}}&
\multicolumn{1}{c|}{\parbox{1.0cm}{[Ni/Fe]}}&
\multicolumn{1}{c|}{\parbox{0.3cm}{$\sigma$}}&
\multicolumn{1}{c|}{\parbox{1.0cm}{[Cu/Fe]}}&
\multicolumn{1}{c|}{\parbox{0.8cm}{$\sigma$}}&
\multicolumn{1}{c|}{\parbox{1.0cm}{[Zn/Fe]}}&
\multicolumn{1}{c|}{\parbox{0.8cm}{$\sigma$}}&
\multicolumn{1}{c|}{\parbox{1.0cm}{[Sr/Fe]}}&
\multicolumn{1}{c|}{\parbox{0.8cm}{$\sigma$}}&
\multicolumn{1}{c|}{\parbox{1.0cm}{[Y/Fe]}}&
\multicolumn{1}{c|}{\parbox{0.8cm}{$\sigma$}}&
\multicolumn{1}{c|}{\parbox{1.0cm}{[Zr/Fe]}}&
\multicolumn{1}{c|}{\parbox{0.8cm}{$\sigma$}}&
\multicolumn{1}{c|}{\parbox{1.0cm}{[Mo/Fe]}}&
\multicolumn{1}{c|}{\parbox{0.8cm}{$\sigma$}}&
\multicolumn{1}{c|}{\parbox{1.0cm}{[Ba/Fe]}}&
\multicolumn{1}{c}{\parbox{0.8cm}{$\sigma$}}\\

\hline
   1    & 33 &34& 35 & 36 & 37  & 38 & 39 & 40 & 41  & 42 & 43 & 44 & 45&46&47&48\\

\hline
NGC 104 &-0.01&0.09&-0.14&0.13&0.22&0.06&0.32&-&0.14&0.23&0.16&0.35&0.51&0.39&0.19&0.25\\
NGC 288 & 0.01&0.07&-0.40&0.03&   -&   -&   -&-&   -&   -&   -&   -&   -&-&0.40&0.10\\
NGC 362 &-0.09&0.03&-0.51&0.02&0.26&0.05&   -&-&0.07&0.11&0.50&0.12&   -&-&0.21&0.06\\
NGC 1261& -   &   -&    -&   -&   -&   -&   -&-&   -&   -&   -&   -&   -&-&-&-\\
NGC 1851& 0.01&0.08&-0.46&0.07&   -&   -&   -&-&0.27&0.15&0.26&0.00&   -&-&0.49&0.01\\
\hline

\end{tabular}
\end{center}

%\label{tproto}
\begin{center}
\begin{tabular}{c|c|c|c|c|c|c|c|c|c|c|c|c|c|c|c|c}

\hline \hline

\multicolumn{1}{c|}{\parbox{1.6cm}{Name}}&
\multicolumn{1}{c|}{\parbox{1.0cm}{[La/Fe]}}&
\multicolumn{1}{c|}{\parbox{0.4cm}{$\sigma$}}&
\multicolumn{1}{c|}{\parbox{1.0cm}{[Ce/Fe]}}&
\multicolumn{1}{c|}{\parbox{0.8cm}{$\sigma$}}&
\multicolumn{1}{c|}{\parbox{1.0cm}{[Nd/Fe]}}&
\multicolumn{1}{c|}{\parbox{0.8cm}{$\sigma$}}&
\multicolumn{1}{c|}{\parbox{1.0cm}{[Eu/Fe]}}&
\multicolumn{1}{c|}{\parbox{0.8cm}{$\sigma$}}&
\multicolumn{1}{c|}{\parbox{1.0cm}{[Dy/Fe]}}&
\multicolumn{1}{c|}{\parbox{0.8cm}{$\sigma$}}&
\multicolumn{1}{c|}{\parbox{1.0cm}{Reference}}\\

\hline
   1    & 49 &50& 51 & 52 & 53  & 54 & 55 & 56 & 57  & 58 & 59 \\

\hline
NGC 104 &0.20&0.10&   -&   -&0.04&0.24&0.42&0.11&0.70&0.10&31,5,37,65,67,82,83,39\\
NGC 288 &   -&   -&   -&   -&   -&   -&0.52&0.12&   -&   -&26,67,18\\
NGC 362 &0.33&0.09&0.14&0.12&0.35&0.10&0.65&0.11&0.68&0.13&26,65,75,18\\
NGC 1261&   -&   -&   -&   -&   -&   -&   -&   -&   -&   -& -\\
NGC 1851&0.38&0.12&0.69&0.20&0.67&0.15&0.71&0.03&0.67&0.12&68,60\\
\hline

\end{tabular}
\end{center}

\end{table}
\end{landscape}

\newpage

\begin{landscape}

\begin{table}[t!]

\caption{%
 Kinematic parameters of globular clusters 
 (Full version of the catalog:  
  http://cdsarc.u-strasbg.fr/viz-bin/qcat?J/AZh/96/267)).}
\bigskip
%\label{tproto}
\begin{center}
\begin{tabular}{c|c|c|c|c|c|c|c|c|c|c|c|c|c}
\hline \hline

\multicolumn{1}{c|}{\parbox{1.6cm}{Name}}&
\multicolumn{1}{c|}{\parbox{1.0cm}{Other name}}&
\multicolumn{1}{c|}{\parbox{0.9cm}{$x$, kpc}}&
\multicolumn{1}{c|}{\parbox{0.9cm}{$y$, kpc}}&
\multicolumn{1}{c|}{\parbox{0.9cm}{$z$, kpc}}&
\multicolumn{1}{c|}{\parbox{0.9cm}{$R_{C}$, kpc}}&
\multicolumn{1}{c|}{\parbox{0.9cm}{$R_{GC}$, kpc}}&
\multicolumn{1}{c|}{\parbox{0.9cm}{$U_{R}$, km\,s$^{-1}$}}&
\multicolumn{1}{c|}{\parbox{0.9cm}{$V_{\Theta}$, km\,s$^{-1}$}}&
\multicolumn{1}{c|}{\parbox{0.9cm}{$U_{Z}$, km\,s$^{-1}$}}&
\multicolumn{1}{c|}{\parbox{0.9cm}{HBR}}&
\multicolumn{1}{c|}{\parbox{0.9cm}{$M_{V}$, mag}}&
\multicolumn{1}{c|}{\parbox{1.6cm}{subsystem}}&
\multicolumn{1}{c}{\parbox{1.2cm}{gen. relation}}\\

\hline
   1  &  2     & 3    &  4   & 5   &  6   & 7    &  8   &9    &10   &11&12&13&14\\
\hline
NGC 104 &47 Tuc& 1.9& -2.6& -3.1& 6.6& 7.4& 55&184&  22&-0.99&-9.42& 2& G\\
NGC 288 &      &-0.1&  0.0& -8.9& 8.6&  12& 24&-74&  52& 0.98&-6.74& 3& -\\
NGC 362 &      & 3.1& -5.1& -6.2& 5.4& 9.4& 71&-35& -68&-0.87&-8.41& 3& -\\
NGC 1261&      & 0.1&-10.0&-12.9& 8.4&18.1&  -&  -&   -&-0.71&-7.81& -& -\\
NGC 1851&      &-4.2& -8.9& -6.9&12.7&16.6&191&142&-102&-0.32&-8.33& 3& -\\

\hline

\end{tabular}
\end{center}

\end{table}
\end{landscape}

%\end{landscape}

\end{document}